\documentclass[preprint,11pt]{aastex}
\usepackage{epsfig}

\newcommand{\be}{\begin{equation}}
\newcommand{\ee}{\end{equation}}
\newcommand{\bt}{\begin{table} \begin{center}}
\newcommand{\et}{\end{center} \end{table}}
\newcommand{\ba}{\begin{eqnarray}}
\newcommand{\ea}{\end{eqnarray}}

\newcommand{\citenp}[1]{\citeauthor{#1}~\citeyear{#1}}

\newcommand{\mt}{\mathit}
\newcommand{\ms}{{\cal M}_\odot}

\def\lesssim{\mathrel{\mathchoice {\vcenter{\offinterlineskip\halign{\hfil
$\displaystyle##$\hfil\cr<\cr\sim\cr}}}
{\vcenter{\offinterlineskip\halign{\hfil$\textstyle##$\hfil\cr
<\cr\sim\cr}}}
{\vcenter{\offinterlineskip\halign{\hfil$\scriptstyle##$\hfil\cr
<\cr\sim\cr}}}
{\vcenter{\offinterlineskip\halign{\hfil$\scriptscriptstyle##$\hfil\cr
<\cr\sim\cr}}}}}

\begin{document}

\title{The porous atmosphere of $\eta$-Carinae}
\author{Nir J. Shaviv}
\affil{Canadian Institute for Theoretical
        Astrophysics, University of Toronto \\ 60 St. George St.,
        Toronto, ON M5S 3H8, Canada}

\begin{abstract}
  
  We analyze the wind generated by the great 20 year long
  super-Eddington outburst of $\eta$-Carinae.  We show that using
  classical stellar atmospheres and winds theory, it is impossible to
  construct a consistent wind model in which a sufficiently {\em
  small} amount of mass, like the one observed, is shed.  One expects
  the super-Eddington luminosity to drive a thick wind with a mass
  loss rate substantially higher than the observed one.  The easiest
  way to resolve the inconsistency is if we alleviate the implicit
  notion that atmospheres are homogeneous.  An inhomogeneous
  atmosphere, or ``porous", allows more radiation to escape while
  exerting a smaller average force.  Consequently, such an atmosphere
  yields a considerably lower mass loss rate for the same total
  luminosity. Moreover, all the applications of the Eddington
  Luminosity as a strict luminosity limit should be revised, or at
  least reanalyzed carefully.

\vskip 0.5cm
\centerline{\em To appear in the Astrophysical Journal Letters}
\vskip 0.5cm

\end{abstract}

\keywords{
  Radiative transfer --- hydrodynamics --- instabilities --- stars:
  atmospheres --- stars: individual ($\eta$ Carinae) }


\section{Introduction}

  $\eta$-Carinae is probably one of the most remarkable stellar object
  to have ever been documented. About 150 years ago, the star began a
  20 year long giant eruption during which it radiated a
  supernova-like energy of roughly $3 \times 10^{49}~ergs$~
  (\citenp{DH97}).  Throughout the eruption it also shed some
  $1-2~\ms$ of material carrying approximately $6\times 10^{48}~ergs$
  as kinetic energy (\citenp{DH97}), while expanding at a velocity of
  $650~km/sec$ (\citenp{HA92}, \citenp{C96}). $\eta$-Carinae can
  therefore serve as a good laboratory for the study of atmospheres
  at extreme luminosity conditions.

  At first glance, it appears that the star shed a large amount of
  material. Indeed, the inferred mass loss rate during the great
  eruption of $\sim 0.1~\ms/yr$ is significantly larger than the mass
  loss rate inferred for the star today ($\lesssim 10^{-3}~\ms/yr$,
  \citenp{DH97} and references therein). However, considering that the
  luminosity during the great eruption is estimated to be
  significantly above the Eddington limit, we shall show that the star
  should have had a much higher mass loss rate. In fact, it should
  have lost during the 20 year eruption more mass than its total mass,
  giving rise to an obvious discrepancy.

  A review of our current knowledge of $\eta$ Car can be found in
  \cite{DH97}. In section \ref{sec:wind} we summarize how a wind
  solution for the star $\eta$ Car should be constructed. Since the
  luminosity is very high, the effects of convection must be taken
  into account. In section \ref{sec:discrepancy} we integrate the wind
  equations to show that no consistent solution for $\eta$ Car exists
  within the possible range of observed parameters. Section
  \ref{sec:bad} is devoted to possible classical solutions to the
  discrepancy, showing that no such possibility exists. In section
  \ref{sec:good}, we show that a porous atmosphere is a simple and
  viable solution to the wind discrepancy.

\section{Solving for the Wind}
\label{sec:wind}

  Since the mass of $\eta$-Car is estimated to be of order
  $100-120~\ms$ (\citenp{DH97}), the average luminosity in the great
  eruption was clearly super-Eddington (of the order of 5 times the
  Eddington limit).  That is to say, the radiative force upwards,
  assuming the smallest possible opacity (for ionized matter) given by
  Thomson scattering, was significantly larger than the gravitational
  pull downwards.  Optically thin winds formally diverge at the
  Eddington limit (e.g., \citenp{Puls} and references therein).
  Consequently, a consistent wind solution requires an optically thick
  wind. We thus look for a wind in which the sonic point (which is the
  point at which the local speed of the outflow equals the speed of
  sound) is below the photosphere.  Moreover, since the duration of
  the eruption is longer than the sound crossing time of the star by
  about a factor of 50, a stationary wind appears to be a good
  approximation.

  In practically all super-sonic wind theories which describe
  super-sonic outflows from an object at rest, a consistent stationary
  solution is obtained only when the net driving force of the wind
  (excluding the pressure gradient) vanishes at the sonic
  point\footnote{ The exception is line driven winds in which the
  force is explicitly a function of $dv/dr$ which is actually an
  approximation to the line transfer equations. If we had written the
  proper radiation transfer equations for this case which only {\em
  implicitly} depend on $dv/dr$, we would have recovered that the
  sonic point coincides with the point at which the total force
  vanishes (cf \citenp{MWM84} \S107).  Moreover, line driven winds are
  important only under optically thin conditions while we describe the
  optically thick part of the wind.  }.  Thus, material experiencing a
  super-Eddington flux necessarily has to be above the sonic point.
  If most of the envelope carries a super Eddington flux, then no
  consistent stationary wind solution can be obtained and in fact, the
  object will evaporate on a dynamical time scale. In most systems
  however, this need not be the case. For example, in very hot systems
  (e.g., hot neutron stars during strong X-ray bursts, \citenp{QP85}),
  the opacity in the deep layers is reduced due the reduced
  Klein-Nishina opacity for Compton scattering at high
  temperatures. Thus, the sonic point in these objects is found where
  the temperature is high enough to reduce the opacity to the point
  where the flux corresponds to the local Eddington limit.

  Another important effect, which should be taken into account, is
  convection.  Deep inside the atmosphere, convection can carry a
  significant part (or almost all) of the energy flux, thus reducing
  the radiative pressure to a sub-Eddington value. In fact, as the
  radiative flux approaches the Eddington limit, convection generally
  arises and carries the lion share of the total energy flux (if it
  can) to keep the system at a sub-Eddington level
  (\citenp{JSO73}). Although the total flux in the entire envelope (or
  almost all of it) can be equivalent to a super-Eddington flux, up to
  some depth below the photosphere, convection carries most of the
  flux so as to reduce the radiative flux alone into a sub-Eddington
  value. A consistent wind solution should therefore, have its sonic
  point at the location where the most efficient convection cannot
  carry enough flux any more.  As we shall soon see, the problem in
  $\eta$-Car is that this point is relatively deep within the
  atmosphere, where the density is so high that the expected mass loss
  is significantly {\it higher} than the observed one.

  To see this in a robust way we integrated numerically the wind
  equations starting from the photosphere inwards. The equations are
  those that describe optically thick spherically symmetric winds
  (\citenp{QP85}; \citenp{Z73}; \citenp{KH94}). The equations of mass
  conservation, momentum conservation and temperature gradient are
\begin{equation}
  4 \pi r^2 \rho v = \dot{M} = {\rm const}
\end{equation}
\begin{equation}
  v {d v\over dr} + {GM\over r^2} +{1\over \rho}{dP_{g} \over dr} -
  {\chi {\cal L}_r \over 4 \pi r^2 c}  =  0
\end{equation}
\begin{equation}
   {d T \over dr} = - {3 \chi \rho {\cal L}_r \over 16 \pi r^2 c a T^3} (1 +
   {2 \over 3 \chi \rho r}),
\end{equation}
with standard notation.  The parenthesized term in the last equation
is a simple approximate interpolation that has the correct asymptotic
limits for optical depths much larger and much smaller than unity
(\citenp{QP85}).  The last equation is the integrated form of the
energy conservation equation. Unlike the aforementioned references, we
specifically include advection by a maximally efficient convection.
Thus, the integrated form of the energy conservation equation becomes
\begin{eqnarray}
    {\cal L}_r + \dot{M}\left({v^2 \over 2} + w - {GM\over
   r}\right) + {\cal L}_{conv} &\equiv& \\ {\cal L}_r -{GM \dot{M}  \over r} +
   {\cal L}_{adv}+{\cal L}_{conv} &=& \Lambda_{tot} = {\cal L}_{obs}+
   {\cal L}_{kin,\infty},
\label{eq:fluxes}
\end{eqnarray}
  were $\Lambda_{tot}$, $\dot{M}$, ${\cal L}_{obs}$ and ${\cal
  L}_{kin,\infty}$ are the total energy output of the star, the wind
  mass loss rate, the observed luminosity at infinity and the kinetic
  energy flux at infinity. On the other hand, ${\cal L}_r, v, w, {\cal
  L}_{conv}, {\cal L}_{adv}$ are respectively, the local radiative
  luminosity, velocity, enthalpy, convective flux and advected flux
  (as internal and kinetic energies).  The expression adopted for
  ${\cal L}_{conv}$ is $4 \pi r^2 u v_{s}$ where $u$ is the internal
  energy per unit volume and $v_s$ is the speed of sound.  By no means
  can convection be more efficient than this expression since highly
  dissipative shocks are unavoidable at higher speeds. It is likely
  that the maximally efficient convection is somewhat less efficient
  than this expression, but this will only aggravate the problem that
  we shall soon expose.  Detailed calculations of the wind were
  carried out.  The calculations include the latest version of the
  OPAL opacities (\citenp{IR96}).  It is found that the total opacity
  below the photosphere has comparable contributions from Thomson
  scattering and absorption processes. This implies that the {\em
  modified} Eddington limit, in which the Thomson opacity is replaced
  by the local total opacity, is somewhat lower than the classical
  Eddington limit.

  Since $T_{\mathit{eff}} \sim 9000^{\circ}$K\footnote{This is the
  typical observed effective temperature for LBV's during outbursts
  (\citenp{HD94}). If the temperature is higher than this value, the
  inferred bolometric magnitude of $\eta$ Car during the eruption
  would be more negative, increasing the Eddington factor. If the
  temperature is lower than $\sim 7000^\circ$K, the opacity at the
  photosphere and outwards rises abruptly (\citenp{D87}), thus
  reducing the modified Eddington limit. In both cases, the
  discrepancy will be aggravated.}, the average luminosity implies a
  photospheric radius of $10^{14}~cm$. Note that since it is a thick
  wind, the exact definition of the photosphere is ambiguous.
  Nevertheless, different definitions do not change the results by
  more than $10-20\%$.  Knowing that the observed mass loss rate is
  roughly $0.1~\ms/yr$ (which gives the observed $2~\ms$ of shed
  material in 20 years, \citenp{DH97}), a specified flow speed at the
  photosphere can be translated to a required density. We can
  therefore integrate the wind equations inwards.  If a consistent
  wind solution can be obtained for some value of the imposed velocity
  in the photosphere (which has to be between $v_s$ and $v_{\infty}$),
  then the integration inwards should reach a sonic point at which the
  total force on the gas vanishes.  This will be attained if the
  convective and advective fluxes can carry a significant amount of
  the total flux so as to reduce the residual radiative flux to a
  sub-Eddington one.

\section{The Discrepancy}
\label{sec:discrepancy}

  We define the luminosity needed to be carried by convection and
  advection in order to bring about the vanishing of the total local
  force as ${\cal L}_{crit}$. If enough energy is advected and
  convected then ${\cal L}_r$ will be reduced to the local modified
  Eddington flux:
\begin{equation}
{\cal L}_{Edd,mod}={4 \pi c G M\over \chi}
\end{equation}
  with $\chi$ the local opacity which can be larger than the Thomson
  opacity.   Thus, from eq.~(\ref{eq:fluxes}), the critical
  advective+convective flux can be written as
\begin{equation}
{\cal L}_{crit} = \Lambda_{tot} + {G M {\dot M}\over r}-{\cal L}_{Edd,mod}.
\end{equation}

\begin{figure}[p]
\centerline{\epsfig{file=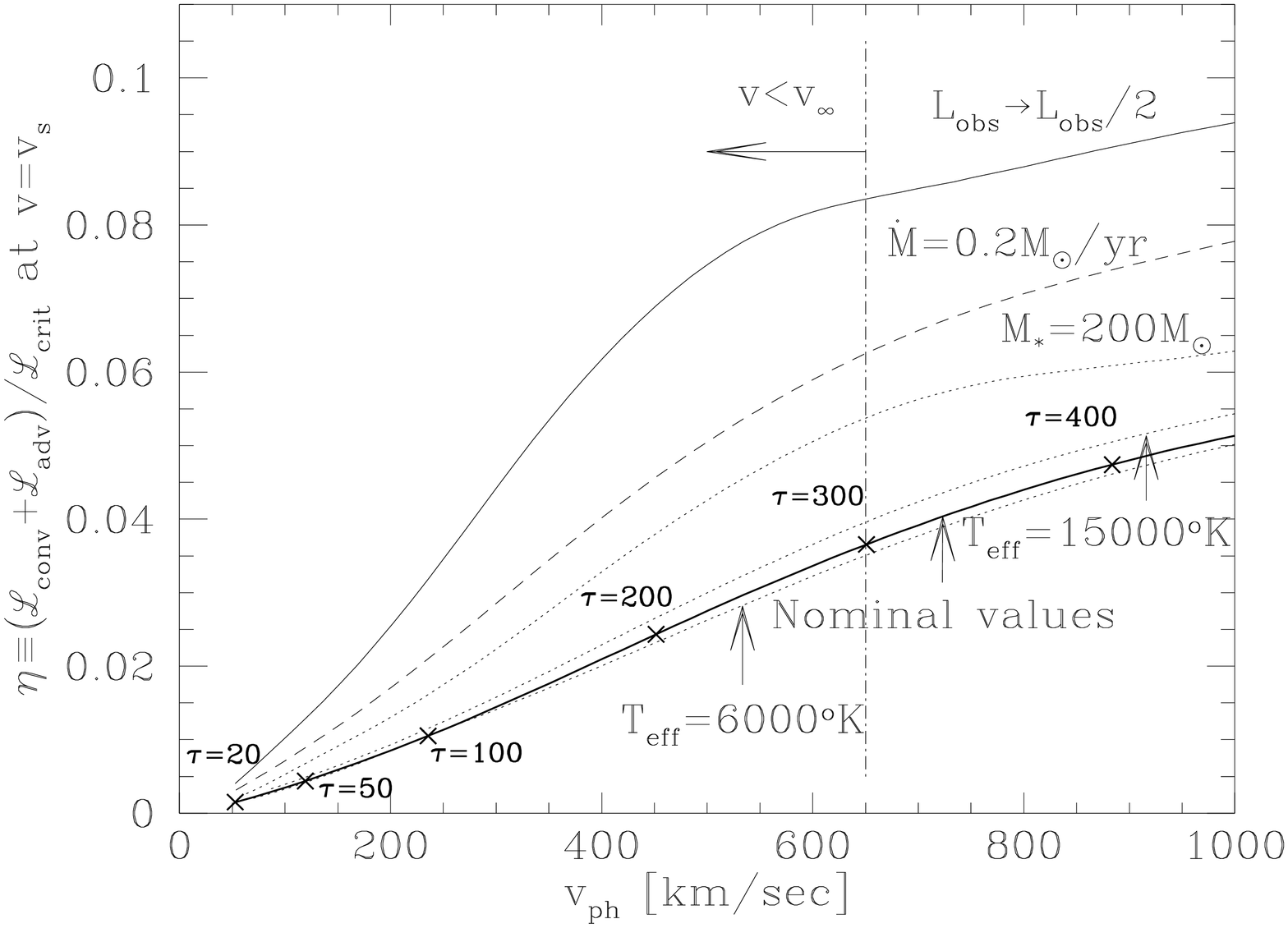,width=4in,angle=-90}}
\caption{
  The fraction $\eta = ({\cal L}_{adv}+ {\cal L}_{conv})/{\cal
  L}_{crit}$ as a function of the photospheric velocity $v_{ph}$.  A
  consistent wind solution requires (a) that the velocity at the
  photosphere $v_{ph}$ satisfy: $v_s \le v_{ph}
 \le v_\infty$, and
  (b) , $\eta(v=v_{s})=1$.  The thick line corresponds to the nominal
  observed and inferred values ($M_\star=100~\ms$,
  $T_{\mathit{eff}}=9000^\circ$K, $\dot{M}=0.1~\ms/yr$, $\int{\cal
  L}_{obs}dt = 3\times 10^{49}~erg$) while the additional lines depict
  the result when the values are changed to their reasonable limits
  (and even beyond). Clearly, no reasonable choice of parameters can
  result with a sonic point that is consistent with a wind solution
  (namely, we always find $\eta(v=v_s)\ll 1$). Basically, the
  discrepancy arises because the mass loss rate observed is too small
  to have the sonic point deep enough in the atmosphere where
  convection can be an efficient mean of energy transport. }
\label{fig_1}
\end{figure}

  Figure \ref{fig_1} shows the fraction $\eta\equiv({\cal
  L}_{adv}+{\cal L}_{conv})/{\cal L}_{crit}$ at the sonic point. A
  consistent solution can be found only if $\eta=1$ at the sonic
  point.

  Inspection of the figure clearly shows that the space of possible
  observed values does not contain a viable and consistent solution.
  This is of course irrespective of whether a solution from the
  photosphere outward can or cannot be obtained. The discrepancy
  arises because a wind corresponding to the observed low mass loss
  rate necessarily has a sonic point that is not deep enough to have
  either convection or advection as an efficient mean of transporting
  energy.  This can be seen from the optical depth at which the sonic
  point is obtained. In all cases, $1 \lesssim \tau < 300$.  However,
  convection is efficient only up to an optical depth of $\tau \sim c/
  v_s \gg 300$ for $p_{rad} \sim p_{gas}$ (\citenp{S00b}), or even
  deeper for larger radiation pressures (i.e., when close to the
  Eddington limit).

\section{Unfeasible Solutions to the Discrepancy}
\label{sec:bad}

  Can the discrepancy be resolve with a classical assumption?  Since
  the discrepancy is rather large, assuming the wind to emerge from an
  angular fraction $f$ from the star does not relax the problem (it
  actually aggravates the problem because more material will be blown
  away from the higher luminosity regions).  Another possibility that
  fails is having a higher velocity in the photosphere than the one
  observed today for the shed material. This might be the case if the
  wind collides with previously ejected slow moving material. Even if
  such material did exist, the necessarily reduced mass loss rate
  inferred from the present day observed momentum aggravates the
  problem.

  The problem is not mitigated if we relax the assumption that the mass
  loss rate and the luminosity are assumed to be constant in time
  throughout the eruption.

  If one wishes to solve the problem using magnetic fields, then a
  solution can be found only if the magnetic energy density at the
  photosphere is significantly larger (by several orders of magnitude)
  then the equipartition value with the gas pressure. This of course
  seems unlikely.

  Another option is to have the distance estimate to $\eta$-Car be
  three times smaller than $2300~pc$. A shorter distance will remove
  $\eta$-Car out of the cluster Tr16 of massive stars inside which it
  is observed and leave it instead roaming the inter galactic arm
  space. Considering the short lifetime of the star, just a few
  million years, this possibility appears as very unlikely.

  The problem can be solved if the mass of the star corresponds to a
  sub-Eddington luminosity. This proposed solution requires $\eta$-Car
  to be at least a $\sim 1000~\ms$ star.  However, this suggestion is
  at variance with much lower estimates (see for instance
  \citenp{DH97} and references therein). Nevertheless, having such a
  massive star is in fact not completely unrealistic and would have
  far reaching consequences if found to be true.

\section{A Viable Solution to the Discrepancy}
\label{sec:good}

  As the title suggests, there is a clear solution to the
  discrepancy. As the results show, the sonic point appears to be
  between the optical depths of $\sim 1$ and $\sim 300$.  The exact
  value cannot be obtained since it requires the integration outward
  from the photosphere, which owing to the relatively inaccurate
  effective temperature and therefore opacity, yields a wide range of
  results. If the {\it mean} radiative force between the point $\eta=1$
  and the above found optical depth, is smaller than classically
  estimated, then a solution to the discrepancy can be found. Such a
  reduction in the mean radiative force is a natural result if the
  atmosphere is inhomogeneous.

  \cite{S98} has shown that in an inhomogeneous atmosphere, the
  effective opacity used to calculate the average force is reduced
  relative to the effective opacity used for the radiation transfer in
  a homogeneous medium.  The effective opacity used for the average
  force should be a volume {\em flux weighted} average of the opacity
  per unit volume\footnote{When the flux is frequency dependant, a
  similar average should be taken in order to find the radiative
  force.  However, one then takes a flux weighted mean over {\em
  frequency space}.}  $\chi_v \equiv \chi \rho$. Namely,
\begin{equation}
  \chi_{\mt{eff}} = {\left< \chi \rho F\right> \over \left< F \right>
  \left<\rho\right>}.
\end{equation}

  The effect is universal and arises in inhomogeneous systems that
  conduct heat or electricity.  Extensive discussions exist in the
  literature under a different terminology (\citenp{I92}). The only
  requirement is therefore, that close to the Eddington limit the star
  develop inhomogeneities.  The transformation from an homogeneous to
  an inhomogeneous atmosphere at luminosities close to but below
  Eddington luminosities, was recently found to take place generically
  even in Thomson scattering atmospheres (\citenp{S99}; \citenp{ST99};
  \citenp{S00}).

\begin{figure}[p]
\centerline{\epsfig{file=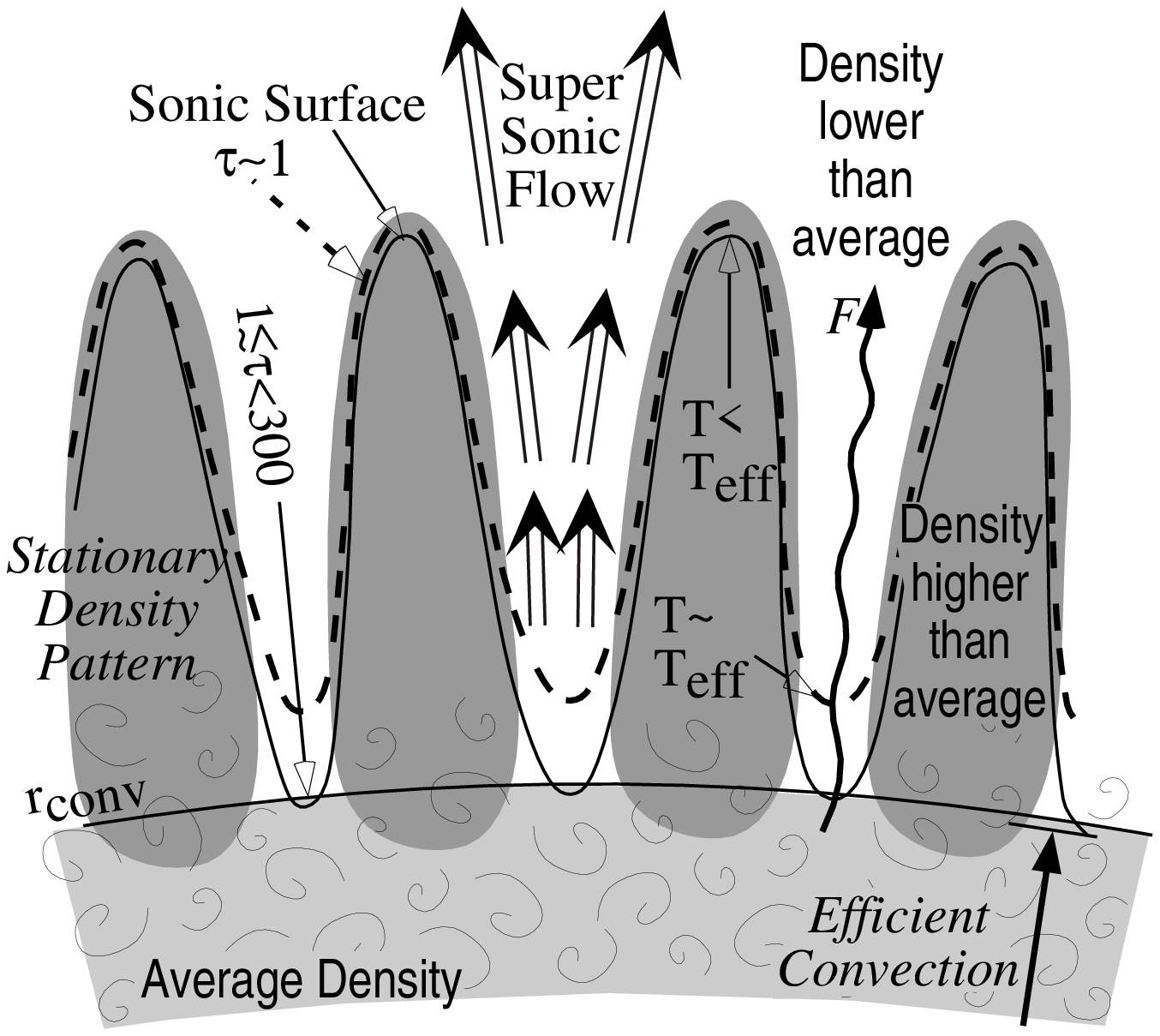,width=3.5in}}
\centerline{\epsfig{file=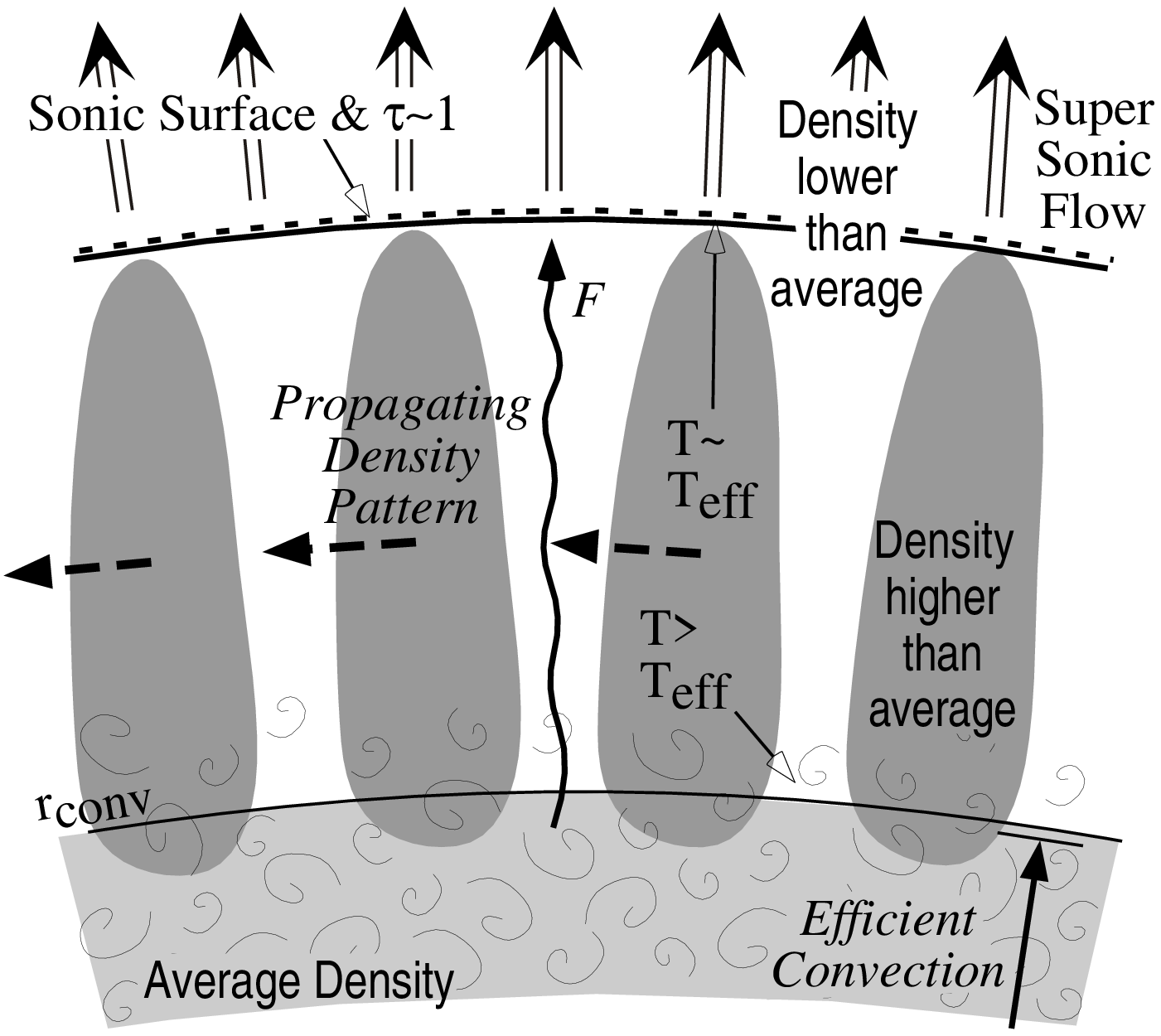,width=3.5in}}
\caption{
 The proposed atmospheric structure of $\eta$ Carinae during its great
 eruption. A homogeneous atmosphere is unstable as a result of two
 generic instabilities that take place even in Thomson atmospheres
 when close to the Eddington limit (\citenp{S00}). The effective
 opacity is therefore reduced (\citenp{S98}) and with it the average
 radiative force. The two panels describe the two types of
 possibilities for having a `porous' atmosphere according to the
 characteristics of the instability that arises. An instability could
 produce a stationary pattern (first panel) if it originates from the
 phase transition instability and a moving pattern if it originates
 from the finite speed of light instability (second panel).  See
 details in the text.}
\label{fig_2}
\end{figure}

  It was found that two different types of instabilities arise
  naturally when the luminosity approaches the Eddington limit
  (\citenp{S00}). One instability is of a phase transition into a {\em
  stationary} nonlinear pattern of ``fingers" that facilitate the
  escape of the radiation.  The second type of instability allows the
  growth of a propagating wave, from which one expects a {\em
  propagating} nonlinear pattern to form.  The two possibilities are
  summarized in figure \ref{fig_2}.  Both instabilities bring about a
  reduction of the average radiative force on the matter and a
  significant reduction of the mass loss rate since the sonic surface
  can sit near (or not much below) the photosphere. In both cases, the
  nonlinear pattern is necessarily expected to form in the region
  between the radius $r_{conv}$ at which $\eta=1$, or in other words,
  that ${\cal L}_{conv} + {\cal L}_{adv}$ is large enough to have
  ${\cal L}_r \lesssim {\cal L}_{Edd,mod}$, and the photosphere. When
  the pattern is stationary, the rarefied regions have a larger than
  Eddington flux and the sonic surface in these regions is near
  $r_{conv}$. On the other hand, if the pattern is propagating, the
  flux may be larger locally than the Eddington limit but the time
  average of the force on a mass element is less than Eddington. Since
  the instability does not occur above the photosphere, it should be
  homogeneous and hence super-Eddington with a super-sonic flow.

  Further analysis of the instabilities is needed to know which
  instability will dominate though it is more likely to be the phase
  transition since it is dynamically more important.

\section{Summary}
\label{sec:summary}

  To summarize, the super-Eddington luminosity emitted by $\eta$-Car
  should have generated a much thicker wind with a sonic point placed
  significantly deeper than what can be directly inferred from the
  observations.  A solution which lives in harmony with observations
  and theoretical modeling is a porous atmosphere, which allows more
  radiation to escape while exerting a smaller average force. It also
  means that the Eddington limit is not as destructive as one would a
  priori think it must be, even in a globally spherically symmetric
  case. Namely, all astrophysical analyses that employ the Eddington
  limit as a strict limit should be reconsidered carefully, even if
  they involve only unmagnetized Thomson scattering material. If
  $\eta$-Carinae could have been super-Eddington for such a long
  duration without ``evaporating'', other systems can display a
  similar behavior.

\end{document}